\documentclass[journal=jacsat,manuscript=article]{achemso}

\usepackage{chemformula} 
\usepackage[T1]{fontenc} 
\usepackage{hyperref}
\hypersetup{
	colorlinks,
	citecolor = blue,
	linkcolor = blue,
	urlcolor  = blue }
\usepackage{booktabs}	

\author{Bal Krishan\textsuperscript{\#}}
\affiliation{Department of Mechanical Engineering,
	Indian Institute of Science, Bengaluru, India}

\author{Preetika Rastogi\textsuperscript{\#}}
\affiliation{Department of Chemical Engineering,
	Indian Institute of Technology, Madras, India}
\altaffiliation{Current Affiliation: Material Sciences Division and The Molecular Foundry, Lawrence Berkeley National Laboratory, Berkeley, California, USA}

\author{D. Chaitanya Kumar Rao}
\affiliation{Department of Aerospace Engineering,
	Indian Institute of Technology, Kanpur, India}

\author{Niket S. Kaisare}
\affiliation{Department of Chemical Engineering,
	Indian Institute of Technology, Madras, India}

\author{Madivala G. Basavaraj}
\affiliation{Department of Chemical Engineering,
	Indian Institute of Technology, Madras, India}
\email{basa@iitm.ac.in}

\author{Saptarshi Basu}
\affiliation{Department of Mechanical Engineering,
	Indian Institute of Science, Bengaluru, India}

\alsoaffiliation{Interdisciplinary Centre for Energy Research,
	Indian Institute of Science, Bengaluru, India}
\email{sbasu@iisc.ac.in}

\title[Evaporation of  water-in-oil microemulsion droplet]
{Evaporation of water-in-oil microemulsion droplet}

\abbreviations{IR, AOT}
\keywords{Droplets, evaporation, microemulsion, shell buckling}

\begin{document}

%
%
%
%
%

\def\thefootnote{\#}\footnotetext{Contributed equally}\def\thefootnote{\arabic{footnote}}

\begin{abstract}
	Emulsion fuels have the potential to reduce both particulate matter and NOx emissions and can potentially improve the efficiency of combustion engines. However, their limited stability remains a critical barrier to practical use as an alternative fuel. In this study, we explore the evaporation behavior of thermodynamically stable water-in-oil microemulsions. The water-in-oil microemulsion droplets prepared from different types of oil were acoustically levitated and heated using a continuous laser at different irradiation intensities. We show that the evaporation characteristics of these microemulsions can be controlled by varying water-to-surfactant molar ratio ($\omega$) and volume fraction of the dispersed phase ($\phi$). The emulsion droplets undergo three distinct stages of evaporation, namely pre-heating, steady evaporation, and unsteady evaporation. During the steady evaporation phase, increasing $\phi$ reduces the evaporation rate for a fixed $\omega$. It is observed that the evaporation of microemulsion is governed by the complex interplay between its constituents and their properties. We propose a parameter ($\eta$) denoting the volume fraction ratio between volatile and non-volatile components, which indicates the cumulative influence of various factors affecting the evaporation process. The evaporation of microemulsions eventually leads to the formation of solid spherical shells, which may undergo buckling. The distinction in the morphology of these shells is explored in detail using SEM imaging.
\end{abstract}

\section{Introduction}\label{sec1:introduction}
	The droplet evaporation is a ubiquitous phenomenon that occurs in many natural and industrial processes. A few common examples are drying of drops in food/chemical industries\cite{samborska_powdered_2024,dantas_innovations_2024}, evaporative cooling of electronic devices\cite{kapilan_comprehensive_2023,xu_spray_2021}, fuel droplet vaporization inside combustion engines \cite{williams_combustion_1973,law_recent_1982,chaitanya_kumar_rao_phenomenology_2020, naka_breakup_2020, pandey_combustion_2017}. The evaporation of drops is also a crucial process in pharmaceutical industries\cite{erbil_evaporation_2012,vehring_pharmaceutical_2008}. The evaporation of liquid droplets has been investigated under distinct configurations, including the sessile mode\cite{hua_evaporation_2002, timm_evaporation_2019, gelderblom_evaporation-driven_2022, korenchenko_sessile_2022, teng_evaporation_2023,thampi2023drying}, where the droplet rests on a solid substrate, and the pendant mode\cite{mandal_internal_2012,chini_understanding_2013,mashayek_dynamics_2001,ranz_evaporation_1951,ranz_evaporation_1952}, where the droplet is suspended from a needle or wire. Droplet evaporation involves a multitude of transport processes (heat, mass, and momentum)\cite{faeth_current_1977, law_recent_1982} and ambient conditions (pressure, temperature, and humidity)\cite{chen_effects_2020}. \citeauthor{brutin_recent_2018} categorically summarised the evaporation of droplets on a substrate and discussed various parameters affecting the evaporation process, such as wettability, thermal effects, and associated flow instabilities. However, drying drops in pendant configuration lead to distinct evaporation characteristics compared to the sessile mode. In many industrial applications, such as spray drying and spray combustion, the droplet remains contactless. Therefore, understanding the evaporation characteristics of an isolated contactless drop becomes crucial.
	\par 
	In recent years, there has been a growing emphasis on multi-component liquids such as emulsion fuels, due to their potential application as alternative fuels for combustion applications \cite{rastogi2024emulsions}. This interest stems from their ability to mitigate harmful emissions (NOx) and enhance engine efficiency\citep{kadota_recent_2002,shinjo_physics_2014, gowrishankar2020investigations}. The study of droplet evaporation is crucial for a comprehensive understanding of their spray and combustion behavior. The addition of water to these fuels can impact the evaporation rate and potentially affect their spray and combustion characteristics \cite{rastogi2024emulsions}. For instance, the addition of water tends to lower the combustion temperature since water exhibits exceptionally high heat capacity, resulting in a significant reduction in NO production\cite{kadota_recent_2002,gopidesi_review_2022}. Furthermore, water-in-oil emulsion fuels can potentially improve the atomization characteristics inside the combustion chamber by undergoing vapor bubble-induced breakup\cite{chaitanya_kumar_rao_phenomenology_2020,priyadarshini_atomization_2024,shinjo_physics_2014,rao_atomization_2020}. While extensive research has been carried out on the atomization characteristics of emulsion fuels, a critical knowledge gap exists regarding their evaporation behavior.
	\par
	\citeauthor{shen_model_2020} reported that upon heating the emulsion droplet, there exists a threshold temperature beyond which the interfacial tension between the water/oil interface increases abruptly. As a result, the water sub-droplets can no longer remain dispersed in oil, and depending on the threshold temperature, the dispersed water droplet can either diffuse towards the surface (eventually vaporize) or undergo coalescence resulting in complete phase separation. A recent study explored the evaporation of sessile water-in-oil emulsion droplets and investigated the capillary clustering of water sub-droplets on the evaporation rate\cite{gopu_distributed_2022}. Similarly, it has been shown that for sessile water-in-cyclohexane emulsion droplets, the evaporation rate can be controlled by varying the water concentration\cite{friberg_evaporation_1992}. Emulsions and their distinct evaporation characteristics are also utilized extensively for making engineered particles\cite{vehring_pharmaceutical_2008,de_souza_lima_drying_2020}. For instance, porous particles can be produced by spray-drying of emulsions. Such particles are useful in pharmaceutical applications such as pulmonary drug delivery systems\cite{dellamary_hollow_2000}.

	While emulsions have diverse applications, in this investigation, we primarily focus on their applicability as an alternative fuel for combustion applications. The emulsions we have discussed so far are called \emph{macro emulsions} (usually referred to as emulsions), and the major challenge with these emulsions is their limited stability. The size of dispersed phase sub-droplets in these emulsions is typically above 1 $\mu\text{m}$. Such systems are milky white and kinetically stable (i.e., eventually undergo phase separation with time)\cite{sharma_introduction_1985,preetika2019kinetic}. The stability of the emulsions depends on parameters such as the size distribution of the dispersed sub-droplets, the volume fraction of water sub-droplets, surfactant properties, and preparation method \citep{califano_experimental_2014,goodarzi_comprehensive_2019}. It is established that the microemulsions are thermodynamically stable, optically transparent, and relatively simple to prepare\cite{hoar_transparent_1943,moulik_structure_1998,domschke_aot_2013,rastogi_diesel_2019}. The microemulsions used in this study exhibit ultra-long stability (up to 180 days\cite{domschke_aot_2013,rastogi_diesel_2019}), making them a viable option as a model alternative fuel for combustion applications. Thermodynamic stability of these emulsions is achieved due to the self-organization of the surfactant molecules at the interface between oil and water sub-droplets, which leads to a decrease in the surface energy of the system.
	\par
	In this work, we investigate the evaporation characteristics of microemulsion droplets in a contactless environment using an infrared laser. Heating with laser offers a significant advantage over conventional droplet heating methods by enabling precise control over the heating rate. To better understand the evaporation process, we employ a non-intrusive method that involves complete isolation of the droplet. This isolation eliminates any external interactions that could potentially influence the evaporation rate. We report that a microemulsion droplet undergoes three distinct stages during the evaporation process, viz., pre-heating (Stage I), steady evaporation (Stage II), and unsteady evaporation (Stage III). Following evaporation, the droplet transforms into a solid residual structure. The global overview of these stages is provided in section \hyperref[sec3-1:global-overview]{3.1}. Subsequently, a detailed understanding of these evaporation process is provided based on microemulsion structure, tuned using parameters $\omega$, the molar ratio; $\phi$, the volume fraction and the nature of the base oil, in sections \hyperref[sec3-2-1:stage-I]{3.2.1}, \hyperref[sec3-2-2:stage-II]{3.2.2}, and \hyperref[sec3-2-3:stage-III]{3.2.3}. An in-depth discussion on shell formation and its dependence on the microemulsion parameters is presented in the section \hyperref[sec3-2-3:stage-III]{3.2.3}. We also propose a parameter $\eta$ referred to as the volume fraction ratio, which is useful to understand the cumulative effect of various factors on the evaporation process and the shell characteristics.

\section{Materials and Experimental Methodology}\label{sec2:material-methods}
	Water-in-oil microemulsions used for droplet evaporation studies throughout this work were prepared using water as the dispersed phase and decane, dodecane, p-xylene, and a surrogate (defined as the mixture of dodecane and p-xylene in the ratio 9:1 by mass) as the continuous oil phase (i. e., base oil). For reference, the properties of the components are tabulated in \autoref{tab1:base-liquid-properties}. A double-tailed ionic surfactant, sodium bis(2-ethylhexyl) sulfosuccinate, commonly referred to as AOT or Aerosol-OT ($\pm 97\%$ purity), was used to stabilize the microemulsions. All the chemicals were procured from Merck and used without any further modification.  High purity freshly de-ionized water (resistivity, 18.2 M$\Omega$-cm) obtained from a laboratory milli-Q system (ELGA, Purelab Option-Q, DV 25).
	
	\begin{table}
		\caption{The properties of liquids used to prepare the microemulsions are presented (at 25 \textcelsius). Listed are the surface tension ($\sigma$), kinematic viscosity ($\mu$), latent heat of vaporisation ($\Delta_vH$), boiling point temperature ($T$), vapour pressure ($p_v$), density ($\rho$) and absorption coefficient ($\alpha$) for infrared electromagnetic waves of wavelength, $\lambda$ = 10.6 $\mu m$. A plot showing temperature dependence of vapour pressure calculated from Antoine equation $\left( \log_{10}(p_v) = A - \dfrac{B}{C + T}\right)$, where A,B and C are constants tabulated in \textcolor{blue}{Table S2} is given in supplementary \textcolor{blue}{Figure S1}.}
		\label{tab1:base-liquid-properties}
		\begin{tabular}{lcccccccc}
			\toprule
			\multicolumn{1}{c}{} & \begin{tabular}[c]{@{}c@{}}$\sigma$\\ (mN/m)\end{tabular} & \begin{tabular}[c]{@{}c@{}}$\mu$\\ (mPa.s)\end{tabular} & \begin{tabular}[c]{@{}c@{}}$\Delta_v H$\\ (kJ/mol)\end{tabular} & \begin{tabular}[c]{@{}c@{}}$T$\\ (\textcelsius)\end{tabular} & \begin{tabular}[c]{@{}c@{}}$p_v$\\ (kPa)\end{tabular} & \begin{tabular}[c]{@{}c@{}}$p_v|_{T = 85\text{ \textcelsius}}$\\ (kPa)\end{tabular} & \begin{tabular}[c]{@{}c@{}}$\rho$\\ ($\text{kg/m}^{3}$)\end{tabular} & \begin{tabular}[c]{@{}c@{}}$\alpha|_{\lambda = 10.6 \mu\text{m}}$\\ ($\text{m}^{-1}$)\end{tabular} \\ \midrule
			Water                & 72.75                                                     & 0.79                                                    & 44.0                                                              & 100                                               & 2.30                                                & 57.05 & 997                                                      & $O(10^5)$                                                                              \\
			Decane               & 23.83                                                     & 0.85                                                    & 51.5                                                            & 174                                               & 0.19                                               & 5.10  & 730                                                      &                                                                                        \\
			Dodecane             & 25.35                                                     & 1.36                                                    & 61.4                                                            & 216                                               & 0.018                                            & 0.95     & 750                                                      &                                                                                        \\
			p-Xylene               & 28.90                                                      & 0.58                                                    & 42.4                                                            & 140                                               & 1.16                                             & 18.86     & 860                                                      &                                                                                        \\
			\bottomrule
			\end{tabular}
		\end{table}

\subsection{Preparation of microemulsions}\label{sec2-1:emulsion-preparation}
	Water-in-oil microemulsions consist of thermodynamically stable dispersion of nanometer-sized water sub-droplets stabilized by a monolayer of surfactant. The characteristics of these microemulsions are characterized based on two parameters: (1) molar ratio of water to AOT (i.e., $\omega$)\cite{kotlarchyk1985structure} and (2) dispersed phase volume fraction (i.e., $\phi$). The $\omega$ and $\phi$ are defined as
	\begin{align}  
		\omega = \frac{[\text{H}_2\text{O}]}{[\text{AOT}]} & 		&\text{and,}& &\phi = 	\frac{V_{\text{H}_2\text{O}} + 	V_{\text{AOT}}}{V_{\text{H}_2\text{O}} + V_{\text{AOT}} + V_{\text{oil}}},
		\label{eqn:phi-omega-relations}
	\end{align}
	where \([\text{H}_2\text{O}]\) and \([\text{AOT}]\) denotes the number of moles of water and AOT, respectively, and \(V\) denotes the volume of different components. The parameters $\omega$ and $\phi$ can be used as handles to tune the size and number density of water sub-droplets in the microemulsions. As evident from \autoref{eqn:phi-omega-relations}, increase in the concentration of water leads to increase in the values of $\omega$ and $\phi$ and is typically used as a handle to tune the size and the number density of the dispersed water sub-droplets\cite{rastogi_diesel_2019, domschke_aot_2013}.
	\par
	For a given value of $\omega$ and $\phi$, the corresponding mass of the continuous oil phase and AOT in the water-in-oil microemulsions can be calculated using \autoref{eqn:phi-omega-relations} by providing the mass of water as an additional input variable. To prepare the microemulsion samples, a pre-determined quantity of AOT is precisely weighed and transferred into pre-cleaned glass vials. Subsequently, an appropriate amount of the oil is added to the vials. The mixture is then shaken gently for a few minutes (depending on the relative quantity of AOT and oil) until the AOT is completely dissolved in the oil phase. Later, a weighed amount of deionized water is added to the oil-AOT mixture and shaken gently until the resulting sample becomes a single-phase optically clear fluid. The samples were stored for 3-4 hours at room temperature (25 \textcelsius) to ensure equilibration and homogeneity before any experiments or characterization.
	\par
	To study the effect of $\omega$ and the nature of the continuous oil phase, microemulsions were prepared using decane as the oil phase at various $\omega$ values: 5, 10, 20, and 40. Furthermore, dodecane microemulsions were formulated at two $\omega$ (= 10 and 20) values, whereas p-xylene microemulsion was prepared at $\omega = 10$. To study the effect of dispersed phase volume fraction, samples were prepared for three different $\phi$ (= 0.1, 0.2, and 0.4) values for all the $\omega$ values. Additionally, a surrogate microemulsion (at $\omega=10, \phi=0.4$) has been formulated, where the oil phase is a mixture of dodecane and xylene in the ratio 9:1.
	\par

	\subsection{Structural characterization of the microemulsions}\label{sec3-0:Nanostructure-of-microemulsion}
	Microemulsions are optically clear, thermodynamically stable dispersions with the dispersed water sub-droplets in the size range of a few nanometers. Such small dispersed structures cannot be resolved using microscopic techniques and therefore, we resorted to scattering techniques. While dynamic light scattering (DLS) was employed to elucidate the structure of decane based microemulsions, small angle x-ray scattering (SAXS) and small angle neutron scattering (SANS) have been used to quantify the size, shape, and interactions for dodecane, p-xylene and surrogate microemulsions reported in this work, prior to evaporation experiments. A thorough characterization of the structure of the microemulsions is necessary and is expected to have a significant influence on their evaporation behavior, as will be shown in the subsequent sections. For a more detailed and systematic characterization of these microemulsions formulated at different values of $\omega$ and $\phi$, we refer to the work of Rastogi et al.\cite{rastogi_diesel_2019,rastogi_modulating_2023}. The wide angle x-ray scattering (WAXS) on selected samples and the SAXS measurements were performed using a Nano-inXider instrument (Xenocs, Sassenge, France) equipped with a microfocus sealed-tube  Cu 30 W/30 $\mu$m X-ray source (Cu k-$\alpha$, $\lambda$ = 1.54 $\text{\AA}$). These small angle scattering experiments were performed with facilities available at the ISIS Neutron and Muon Source of the Rutherford Appleton Laboratory (Didcot, UK).
	\par
\begin{figure}[ht!]
		{\includegraphics[width=1\textwidth]{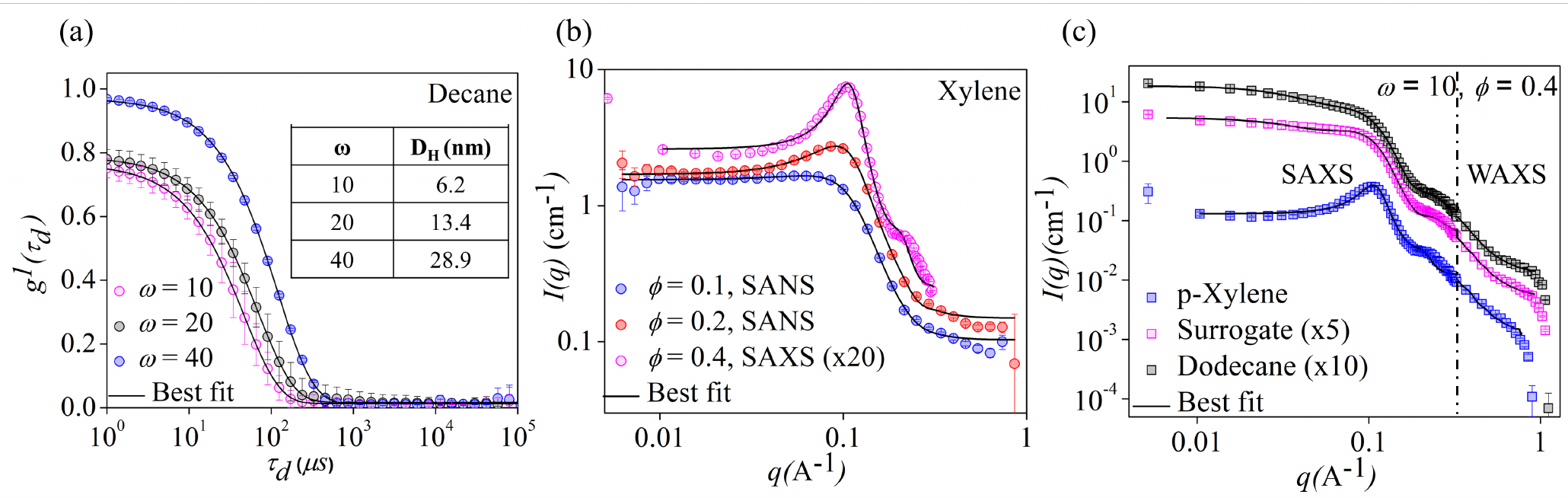}}
		\caption{Structural characterization of water-in-oil microemulsions used for evaporation studies reported in this work. (a) The electric field autocorrelation function ($g^1(\tau)$) versus delay time ($\tau_d$) obtained from dynamic light scattering at 25 $^\circ$C. The effective diffusivity ($D_{eff}$) of the water drops obtained by a fit to the exponential decay of the autocorrelation function together with the Stokes-Einstein relation is used to quantify the size of water sub-droplets in decane oil at $\omega = 10, 20$ and $40$ at $\phi = 0.1$. The data elucidates the effect of an increase in $\omega$ on the size of microemulsion droplets. The hydrodynamic diameter ($D_\text{H}$) of dispersed water nanodroplets is tabulated in the inset. (b) $I(q)$ vs $q$ acquired at 20 \textcelsius from SANS ($\phi = 0.1$ and $0.2$) and SAXS ($\phi = 0.4$) experiments at $\omega = 10$ demonstrating the effect of increase in $\phi$ for the case of water-in-xylene microemulsions. The SANS data is adapted from our previous work\cite{rastogi2023investigation} with permission. SAXS data is scaled by a factor of 20 for clarity. (c) SAXS and WAXS data acquired at 20 \textcelsius for a fixed composition ($\omega = 10$ and $\phi = 0.4$) for water-in-oil microemulsions with different oil phases highlighting the role of chemical nature of the oil phase on the nanostructure of droplets. SAXS and WAXS data for surrogate and dodecane microemulsions are scaled by 5 and 10 respectively. Symbols in all the plots are the experimental data whereas continuous black lines are the corresponding fits.}
		\label{fig:structure}
	\end{figure}

	
	For decane microemulsion, the effective diffusivity ($\mathcal{D}_\text{eff}$) of the water sub-droplets decreases with the increase in the values of both $\omega$ and $\phi$. Furthermore, the hydrodynamic diameter ($D_\text{H}$) of the water sub-droplets is also reported and was calculated based on the Stokes-Einstein relation given as
	\begin{equation}
		D_\text{H} = \frac{kT}{3\pi \mu \mathcal{D}_\text{eff}}
		\label{eqn:Stokes-Einstein-relation}
	\end{equation}
	where, $k$ is the Boltzmann constant, $T$ is the absolute temperature, and $\mu$ is the solvent viscosity. The water sub-droplet diameter for water-in-decane is found to be $D_\text{H} \sim 6.2 \text{ nm}$ for $\omega=10$ and $\phi=0.1$. The viscosity $(\mu)$ of the microemulsion prepared was measured, and it was found that the viscosity increased with the increase in the value $\phi$. \autoref{fig:structure}(a) elucidates the effect of molar ratio, $\omega$ on the size of the water sub-droplets dispersed in a continuous decane phase. The symbols in the plot correspond to the electric field autocorrelation function ($g^1(\tau_d)$) as a function of delay time ($\tau_d$) obtained from dynamic light scattering analysis of decane microemulsions for $\omega = 10, 20$ and $40$ at $\phi = 0.1$. A single exponential decay of the $g^1(\tau_d)$ versus $\tau_d$ curve (black continuous lines) indicates the presence of spherical droplets of water dispersed in the decane phase. The shift in the $g^1(\tau_d)$ observed with an increase in the $\omega$, suggests an increase in the size of the water nanodroplets. For the data shown in \autoref{fig:structure}(a), the hydrodynamic diameter ($D_\text{H}$) of water sub-droplets is estimated using the fitting procedure described earlier  \cite{rastogi_diesel_2019} and the Stokes-Einstein equation (Equation \ref{eqn:Stokes-Einstein-relation}) and is shown in the inset of \autoref{fig:structure}(a).
	\autoref{fig:structure}(b), The scattering intensity, $I(q)$ versus scattering vector, $q$, recorded from neutron ($\phi = 0.1$ and $0.2$) and x-ray scattering ($\phi = 0.4$)  captures the effect of change in volume fraction, $\phi$ on the structure of xylene microemulsion at $\omega = 10$. Since the estimated size of the dispersed water sub-droplets in p-xylene is near the lower limit of the DLS instrument, the use of small angle neutron and X-ray scattering techniques was crucial to gain insight into the structure of these microemulsions. The $I(q)$ versus $q$ data for xylene microemulsions at $\omega = 10$ and $\phi = 0.1$ and $0.2$ is adapted from \citeauthor{rastogi2023investigation}. The water sub-droplets dispersed in xylene are found to be spherical in shape and had a core diameter of 2.70 ± 0.44 for $\phi = 0.1$ and 2.92 ± 0.22 nm for $\phi = 0.2$. More details of the structural analysis can be found elsewhere\cite{rastogi2023investigation}. Due to the limited availability of the beamtime on the SANS beamline, data for $\omega = 10$ and $\phi = 0.4$ was recorded on the SAXS instrument and is scaled by a factor of 20 in \autoref{fig:structure}(b) for depiction alongside data from neutron scattering. The experimental details and the fitting procedure for SANS and SAXS data is illustrated previously \cite{rastogi2023investigation, rastogi_modulating_2023}. For completeness, experimental data from small and wide angle X-ray scattering (SAXS and WAXS) is acquired at $\omega = 10$ and $\phi = 0.4$ for three different oil phases -- p-xylene, dodecane, and surrogate and is shown as symbols in \autoref{fig:structure}(c). The plot shows the effect of the oil phase on the nanostructure of the dispersed water sub-droplets at the same composition. SAXS and WAXS data for surrogate and dodecane microemulsions is shifted by a factor of 5 and 10 respectively for clarity of presentation. The black continuous lines are the structure factor that fits to the combined SAXS and WAXS data (symbols). The water sub-droplets in xylene microemulsions are modeled as spherical objects with sticky hardsphere type interactions. The dispersed species in dodecane and surrogate microemulsions are found to be cylindrical in shape. Therefore, a structure factor for cylinders interacting via sticky interactions is used for the analysis of scattering data from these samples. The fit parameters are tabulated in \autoref{tab2:SAXSFitParams}, which lists the parameter $\chi^2$ which denotes the goodness of the fits, diameter (D) of the spherical droplets in xylene microemulsions, length (L) and diameter (D) of the cylinder for the case of surrogate and dodecane microemulsions, polydispersity index for variation in diameter ($PDI-D$) and length of the cylindrical droplets ($PDI-L$).
	\begin{table}
		\caption{SAXS fit parameters for the data shown in \autoref{fig:structure}(c) for microemulsions of different oils at $\omega = 10$ and $\phi = 0.4$}
		\label{tab2:SAXSFitParams}
		\begin{tabular}{lccccccc}
			\toprule
			Oil Phase & $\chi^2$ & Diameter, D (nm) & $PDI-D$ & Length, L (nm) & $PDI-L$ \\
			\hline
			p-xylene &  29.1 & 2.5 & 0.45 & - & - \\
			Dodecane &  9.8 & 4.2 & 0.17 & 8.9 & 0.19 \\
			Surrogate &  24.8 & 4.1 & 0.15 & 9.5 & 0.07 \\
			\bottomrule
		\end{tabular}
	\end{table}
	
	\subsection{Experimental details and post processing}\label{sec2-2:experimental-methodology}
\begin{figure}[ht]
		\centering
		\includegraphics[width=0.95\textwidth]{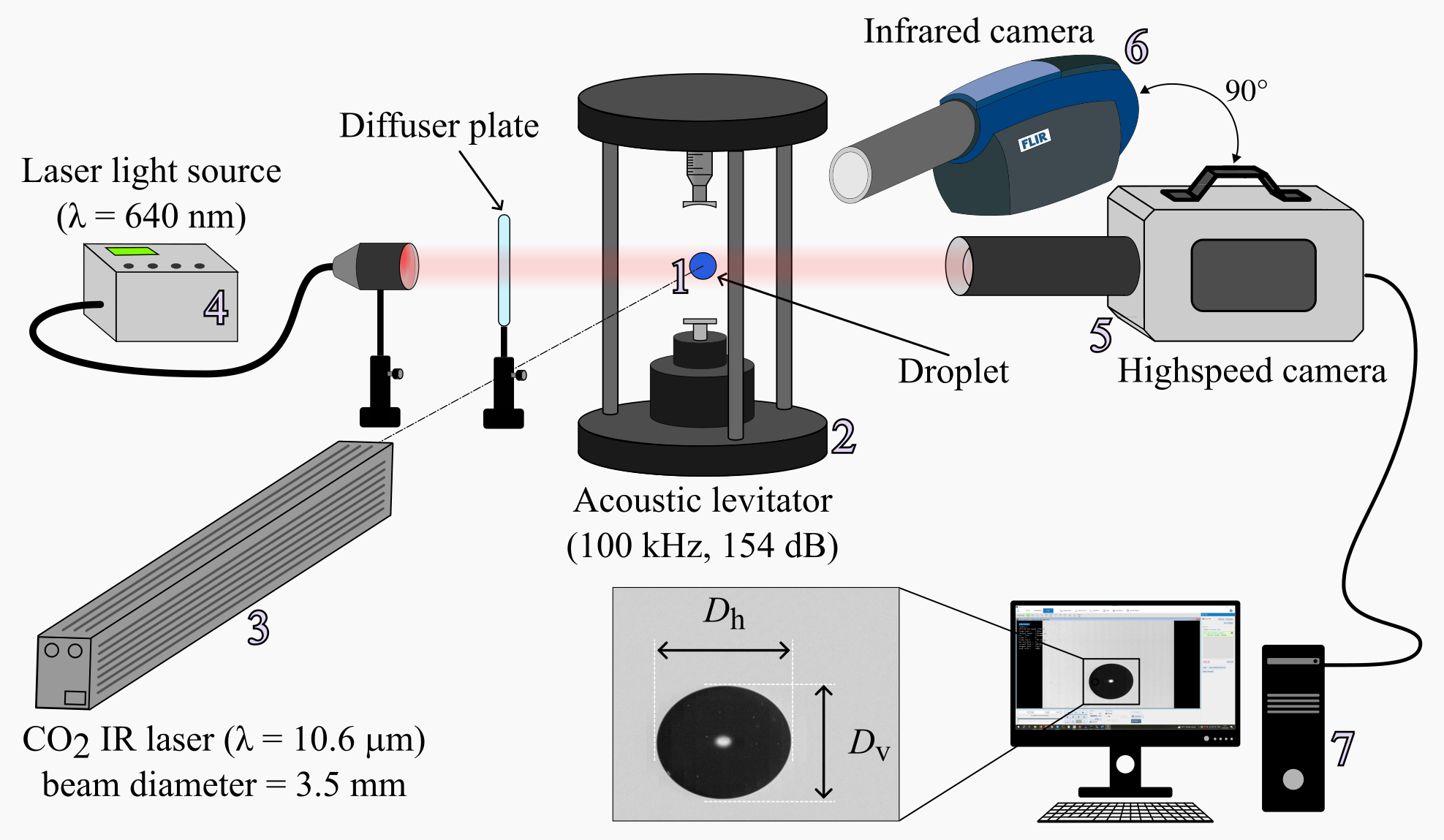}
		\caption{Schematic of the experimental setup showing a droplet (1), levitated using an acoustic levitator (2) and heated using an infrared laser (3). A shadow of the droplet is created using a laser light source (4) upon the censor of a high-speed camera (5) to capture the shadowgraphy images. IR thermography images are recorded using the IR camera (6). The recorded data was stored in the computer (7) for further processing.}
		\label{fig1:experiemtal-setup}
	\end{figure}

	The schematic of the experimental setup is shown in \autoref{fig1:experiemtal-setup}. A droplet of diameter $D_0 \sim 850\pm 50 $ micron is suspended using a single-axis acoustic levitator (Tec5) with a frequency of 100 kHz at standard ambient conditions. The droplet is heated externally at irradiation intensities $I^* = 0.05$, and $0.1$, where $I^* = I/I_{max}$ and $I_{max} = 1.04\text{ MW}/\text{m}^2$ with a 3.5 mm beam diameter continuous \(CO_2\) infrared (IR) laser (Synrad 48, wavelength $\lambda \sim 10.6  \ \mu\text{m}$, and maximum power (\(P_\text{max}= 10  \text{W}\)). Droplet evaporation dynamics are recorded using a high-speed camera (Photron SA5) with a high-speed laser light source (CAVILUX\(^\text{\textregistered}\) Smart UHS, 640 nm). 
	A recording rate of 50-125 frames per second (fps) has been used because of the large time scale associated, along with the spatial resolution of 5.2 $\mu\text{m}$ per pixel. An infrared camera is used for droplet surface temperature measurement at 170 frames per second. At the minimum, six repetitions for each sample were conducted to ensure the reproducibility of data and capture the probability of occurrence of the phenomenon. The droplet images are analyzed and processed using Photron's proprietary software PFV4(x64) and ImageJ (version 1.53t). For data plotting and figure preparation,  Python 3.10 and Inkscape 1.1.2 have been used, respectively. The equivalent diameter of the droplet is calculated as \(D_\text{eq} = \sqrt{D_\text{h}D_\text{v}}\), where \(D_\text{h}\) and \(D_\text{v}\) represent the maximum horizontal and vertical length scales (i.e., diameter) of the droplet as shown in the \autoref{fig1:experiemtal-setup}. The uncertainty associated with the calculated droplet diameter and temperature is less than or equal to $\pm6\%$ and  $\pm7.5\%$, respectively.
	
\section{Results and discussion}\label{sec3:results}


	\subsection{A global overview}\label{sec3-1:global-overview}
	A microemulsion droplet undergoes various changes throughout its lifetime when it comes in contact with the infrared (IR) laser. It is important to note that a liquid droplet may respond to electromagnetic irradiation in many different ways, exhibiting various phenomena such as plasma initiation and breakup\cite{Rao_Laser_induced_2022}, spontaneous nucleation\cite{Park_Laser_1989} and vaporization\cite{Armstrong_aerosol_1984}. Depending on the irradiation intensity, different phenomena have been observed in the past\cite{Park_Laser_1989}. This study investigates the vaporization behavior of microemulsion droplets, which usually requires significantly low irradiation intensity depending on the absorption properties of the liquid. The evaporation of the microemulsion droplet can be divided into three distinct stages primarily characterized by the microemulsion composition($\phi$, $\omega$) and irradiation intensity. Soon after the droplet is exposed to the irradiation, it starts to heat up, and its temperature increases until it reaches a constant value. The temperature rise causes the droplet to undergo evaporation, as a result, the droplet loses its mass and the diameter of the droplet reduces marginally. At the end of evaporation, the formation of a spherical solid residue is observed. As time progresses, this spherical residue undergoes buckling due to the acoustic pressure and stresses generated during the course of evaporation. The formation of the shell is a unique phenomenon observed for microemulsions, primarily due to the \emph{non-volatile} nature of the surfactant (AOT) used in this work. A detailed discussion about different stages of evaporation is presented in the next section \hyperref[sec3-2:evaporation-of-microemulsion]{3.2}.	
	\par 
	\begin{figure}[ht]
		{\includegraphics[width=0.75\textwidth]{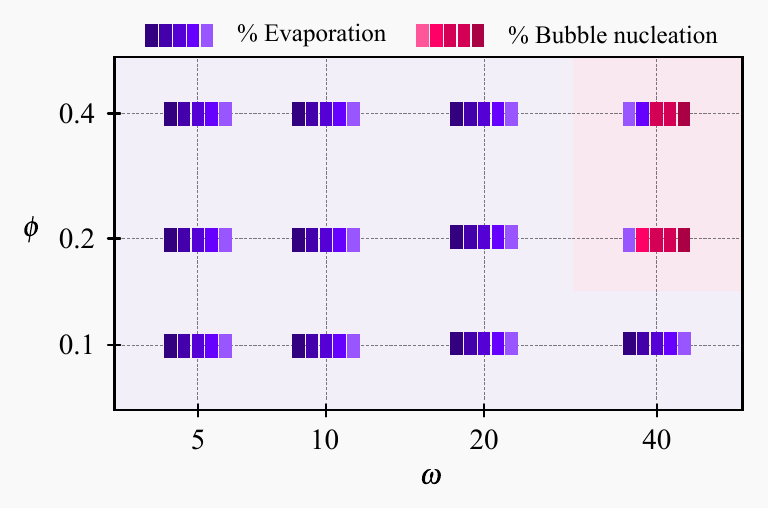}}
		\caption{Probability of occurrence of undisrupted evaporation and bubble nucleation during the heating (at $I^\ast=0.1$) of decane microemulsion formulated at different $\phi$, and $\omega$. Each colored block represents a 20 \% probability of occurrence.}
		\label{fig2:overview}
	\end{figure}
	Primarily, droplet evaporation phenomenon has been observed for all the cases at low irradiation intensity ($I^\ast = 0.05$); however, increasing the irradiation intensity to $I^\ast = 0.1$ results in the nucleation of a vapor bubble inside the droplet for decane and xylene based microemulsions (only for higher $\phi$ and $\omega$ values). These vapor bubbles expand and grow, eventually undergoing breakup/collapse. Typically, the bubble nucleation depends on various factors, such as degree of superheat, IR absorbance, heat of vaporization, and number of active nucleation sites. It is important to note that at $I^\ast = 0.05$, we do not observe bubble nucleation across all the samples (i.e., microemulsions with a particular value of $\omega$ and $\phi$) and trials (i.e., repetition of experiment for a sample) explored in this study (essentially undisruptive evaporation occurs at $I^\ast = 0.05$). However, at $I^\ast = 0.1$, most samples exhibit undisruptive evaporation, except for a decane and xylene microemulsion having higher $\omega$ and $\phi$ value.  It is to be pointed out that the vapor bubble nucleation at $I^\ast = 0.1$ was observed only for a few repetitions in xylene and decane microemulsions, which is expected since the nucleation process is a stochastic phenomenon\cite{mikami_experimental_2002}. To identify the most probable behavior among evaporation/nucleation, we have calculated their probability of occurrence across all the trials. \autoref{fig2:overview} shows the probability of occurrence of evaporation and/or bubble nucleation for decane microemulsion, where each colored block represents the 20\% probability of occurrence.  The dominant phenomena have been decided based on the probability of occurrence. For example, the decane microemulsion at $\omega = 40$ and $\phi = 0.2$ shows an 80\% probability of bubble nucleation compared to 20\% of undisruptive evaporation.
	\par
	To study the effect of surfactant on droplet evaporation, experiments were conducted using mixtures of oil and AOT at different AOT concentrations without adding water. Additionally, for completeness of the study, droplet evaporation experiments were also conducted for pure liquids.
	
	\subsection{Evaporation of microemulsion droplet}\label{sec3-2:evaporation-of-microemulsion}
\begin{figure}
		{\includegraphics[width=0.9\textwidth]{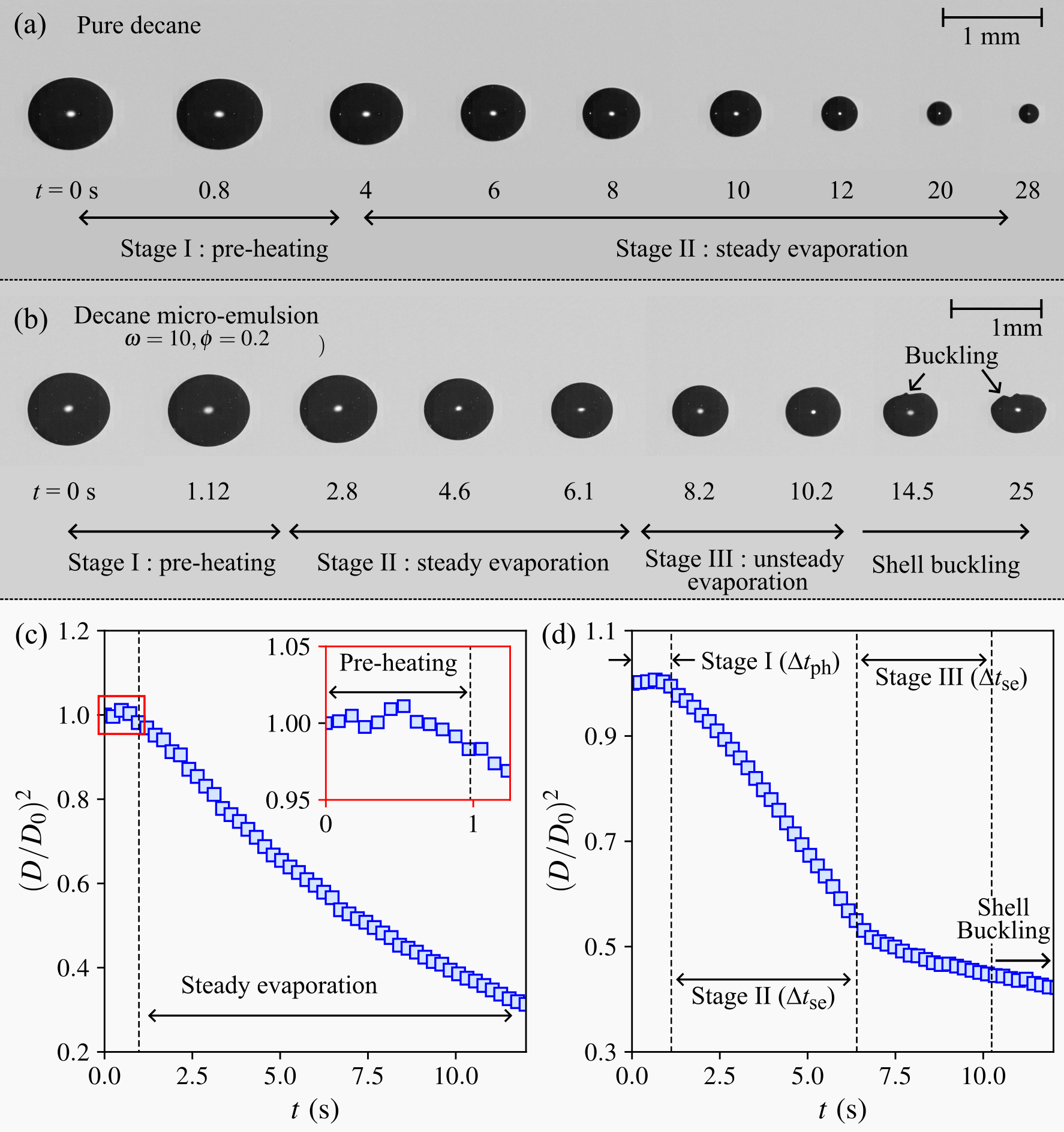}}
		\caption{(a) Time series images of pure decane evaporation depicting different stages of the droplet evaporation for the pure liquid evaporation. (b) Also shows different phases of droplet evaporation for a decane microemulsion ($\omega = 10, \ \phi = 0.2$) droplet, such droplet, after the end of evaporation, forms a shell of non-volatile surfactant (AOT), which undergo buckling under the acoustic pressure. (c) and (d) show the change in the non-dimensional surface area of the droplet with time representing different stages involved during the evaporation of pure decane and decane microemulsion ($\omega = 10, \ \phi = 0.2$) droplets, respectively.  }
		\label{fig3:pure_vs_me_droplet_evaporation}
	\end{figure}

	Typically, the evaporation of a single component liquid droplet occurs in two stages upon interaction with an IR laser. \emph{Stage I} is the preheating Stage, where the irradiation energy increases the temperature of the droplet, with minimal change in its diameter. \emph{Stage II} corresponds to the steady evaporation stage where the droplet diameter reduces steadily (see \autoref{fig3:pure_vs_me_droplet_evaporation}(a), and (c)) with constant surface temperature \cite{chaitanya_kumar_rao_phenomenology_2020, pandey_how_2018}. During the pre-heating phase, the droplet absorbs the irradiation as sensible heat, which increases its temperature. As a result, the droplet's surface temperature also rises until it becomes constant. Following this, the droplet starts to evaporate, and the irradiation energy absorbed by the droplet during this phase is utilized to vaporize liquid from the droplet surface in the form of latent heat. Consequently, due to the mass loss from the droplet surface, the pure liquid droplet undergoes near-complete evaporation, reaching a final diameter significantly smaller than its initial size (${D(t)}/{D_0} \sim 0.1\pm0.05$). Once the droplet reaches this size, it becomes undetectable within the limitations of the current experimental setup, either due to visualization limitations or loss of stable levitation in the acoustic levitator.
	\par
	The fundamental difference between a pure liquid and a microemulsion droplet is that the pure liquid consists of a single component whereas microemulsions are multi-component systems, which typically contain a base oil (i.e., continuous phase), water (i.e., dispersed phase), and a stabilizer or surfactant. In this study, we use microemulsions (water-AOT-oil) droplet, which has nanometer-sized water sub-droplets dispersed into the continuous phase (oil), where a monolayer of surfactant (AOT) stabilizes the interface between the water and oil, as shown in the Figure 5(a). The microemulsions are a ternary system and quaternary for the surrogate microemulsion (water-AOT-dodecane-xylene) and their evaporation characteristics depend on the molar ratio of water to AOT ($\omega$), and dispersed phase volume fraction ($\phi$)\cite{rastogi_modulating_2023,domschke_aot_2013}.
	\par
	The evaporation of a microemulsion droplet has been observed to occur in three different stages, where the first two stages (namely preheating and steady evaporation) are similar to that for single-component droplet evaporation. An additional stage (Stage III) referred to as \emph{unsteady evaporation} stage is observed, where the droplet evaporation rate continuously decreases until the evaporation completely stops, which is attributed to the non-volatile nature of the surfactant AOT. Upon heating, AOT molecules undergo agglomeration as the evaporation proceeds, and their concentration at the droplet surface increases, leading to the formation of a shell at the end of evaporation. During evaporation, the time scales associated with different stages are affected by the components and compositional parameters ($\omega$ and $\phi$) of the microemulsion and the laser irradiance. The droplet evaporation time (i.e., lifetime) (\(\tau\)) of a microemulsion droplet, defined as the time period between the start of laser irradiation on the droplet and the end of evaporation, is given as:
	\begin{equation}
		\tau = \Delta t_\text{ph} + \Delta t_\text{se} + \Delta t_\text{ue}
	\end{equation}	
	where $\Delta t_\text{ph}, \Delta t_\text{se}$, and $ \Delta t_\text{ue}$ denote the preheating, steady,  and unsteady evaporation time periods, respectively. The quantitative measures associated with droplet evaporation, such as evaporation rate and time scales, have been calculated from the experimental data and discussed in subsequent sections.
	
		\subsubsection{Stage I: Pre-heating }\label{sec3-2-1:stage-I}
\begin{figure}
			{\includegraphics[width=0.9\textwidth]{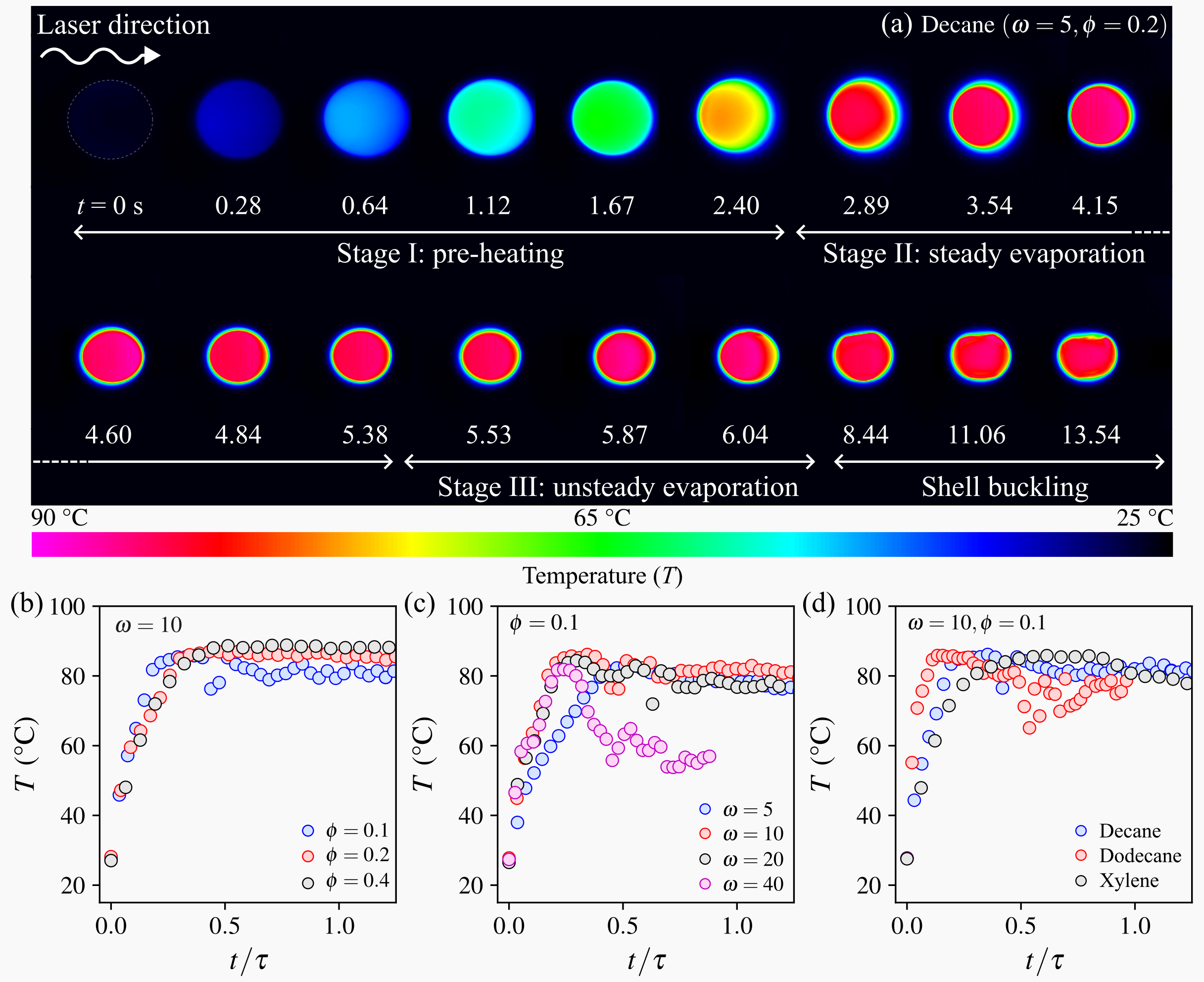}}
			\caption{(a) Infrared images showing the droplet surface temperature distribution during different stages of evaporation for decane microemulsion ($\omega = 5, \phi=0.2$) droplet. Comparing the evolution of droplet surface temperature for different values of (b) $\phi$ at $\omega = 5 $ and at different (c) $\omega$ values at $\phi = 0.1$ for decane microemulsion at a laser irradiation intensity of $I^* = 0.05$. (d) showing droplet surface temperature variation of different oil phase for $\omega = 10, \phi = 0.1 $ at $I^* = 0.05$.  }
			\label{fig4:temperature_plots_and_contours}
		\end{figure}

		As discussed earlier, during the first stage (pre-heating), the laser irradiation increases the temperature of the droplet. It is important to note that droplet heating exhibits directional dependence due to the unidirectional irradiation from the laser. Consequently, the droplet experiences non-uniform heating initially. However, for an acoustically levitated droplet, the rotation\cite{saha_velocity_2012} and internal flow \cite{hasegawa_microlayered_2016,yamamoto_internal_2008} facilitates the transport of energy throughout the droplet. This eventually leads to a more uniform temperature distribution within the droplet, as shown in the \autoref{fig4:temperature_plots_and_contours}(a).
		Following laser activation, the temperature of the droplet exhibits a monotonic increase. This initial heating phase is referred to as the pre-heating period, characterized by a continuous rise in temperature as illustrated in \autoref{fig4:temperature_plots_and_contours}. During the pre-heating phase, the irradiation from the laser is used to increase the internal energy of the droplet, which translates to a rise in its surface temperature until it becomes constant (wet-bulb temperature\cite{raju_controlling_2018,pandey_how_2018}). In this regime, the droplet absorbs energy from the laser as sensible heat, and loss of energy to the surrounding is negligible due to minimal vaporization and conduction (a detailed discussion is presented in section\hyperref[sec3-2-2:stage-II]{3.2.2}). Later, the dominant heat transfer mechanism transitions from sensible heating to latent heat transfer, and this shift in heat transfer dynamics manifests as a plateau in the droplet surface temperature profile. The droplet size does not change significantly during the pre-heating phase; however, the droplet undergoes volumetric expansion, resulting in a marginal increase in the size ($\Delta D/D_0 \leq 1\%\pm 0.5$) of the droplet before the beginning of Stage II. \autoref{fig4:temperature_plots_and_contours}(a) shows the infrared images of the droplet surface temperature distribution during the evaporation of water-in-decane microemulsion ($\omega=5, \phi=0.2$). It is observed that the droplet starts to heat-up from the same direction as the laser, indicating the directional heating of the droplet. The evolution of maximum surface temperature is plotted in the \autoref{fig4:temperature_plots_and_contours}(b)-(d) for different microemulsions.  \autoref{fig4:temperature_plots_and_contours}(b) shows that the temperature starts to increase at similar rate and reaches a steady state temperature for the water-in-decane microemulsions at $\omega = 10$ and $\phi = 0.1, 0.2$ and $0.4$. However, a slight variation in the heating rate can be seen for microemulsions with different base oils, (see \autoref{fig4:temperature_plots_and_contours}(d)), which translates to a variation in their heating time. The preheating time ($\Delta t_\text{ph}$) can be quantified through two independent analyses: the evolution of droplet diameters through high-speed images and the evolution of surface temperature through thermal imaging. The preheating time ($\Delta t_\text{ph}$) can be determined from the temperature profile by identifying the time interval between the initiation of laser irradiation and the point where the temperature reaches a steady-state.
		
		\par
		The preheating time is anticipated to be influenced by the microemulsion parameters ($\omega$, $\phi$) and the properties of constituents (such as IR absorption coefficient, heat capacity, viscosity, and volatility). From \autoref{fig4:temperature_plots_and_contours}(b), it is observed that an emulsion droplet with higher $\phi$ exhibits higher preheating time at a particular $\omega$, however, the difference is minimal. At low heating rate ($I^\ast = 0.05$), the pre-heating time ($\Delta t_\text{ph}$) is $\sim 1.72 \pm 0.53$ seconds, however at higher heating ($I^\ast = 0.1$), this time reduces to $\sim 0.76 \pm 0.35$ seconds. The water-in-decane microemulsion droplet attains a surface temperature ($T$) of $\sim 89 \pm 2$ {\textcelsius}  for $\phi = 0.4$, and it is reduced to $85 \pm 2$ {\textcelsius} for a lower $\phi$ (0.1) at $I^* = 0.05$. A higher temperature for a higher $\phi$ is expected due to the presence of larger number of dispersed sub-droplets. For $\phi=0.1$, for different $\omega$ values, the droplet surface temperature ranges between $60-90 $ {\textcelsius}.  The surface temperatures of microemulsions containing different base oils have also been analyzed; the results show that for all the emulsion droplets, the surface temperature ranges between $80 - 85$ {\textcelsius}.
		
		\subsubsection{Stage II: Steady evaporation}\label{sec3-2-2:stage-II}
		During steady evaporation stage, the droplet undergoes a well-defined, steady-state evaporation phenomenon \cite{pandey_how_2018,chaitanya_kumar_rao_phenomenology_2020,mehrizi_evaporation_2022}, which is characterized by a constant rate of decrease in droplet diameter, as observed in \autoref{fig3:pure_vs_me_droplet_evaporation}.The evaporation phenomenon can be attributed to a balance between the energy supplied by the laser irradiation and the energy required for vaporization of the droplet, considering a negligible amount of the energy is lost to the surroundings via conduction\cite{Park_Laser_1989}. As a result, a cloud of vapor surrounding the droplet surface is created, which then diffuses away due to a concentration gradient.
		\begin{figure}[ht]
			{\includegraphics[width=0.85\textwidth]{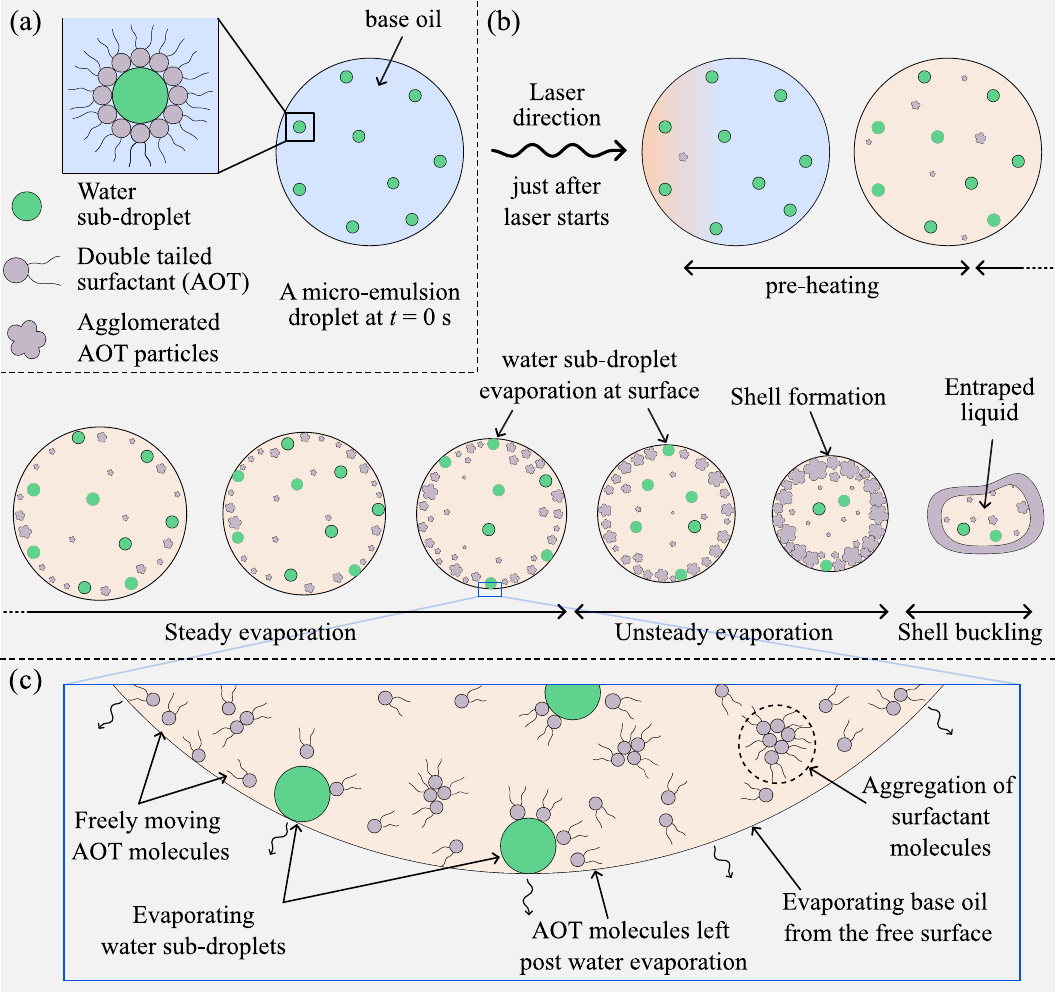}}
			\caption{(a) Schematic showing a microemulsion droplet and its constituents. (b) A microemulsion droplet interacting with an infrared laser, showing different stages of droplet evaporation, shell formation, and buckling. (c) A magnified image (exaggerated) of microemulsion droplet interface showing different phenomena involved during evaporation}
			\label{fig5:schematic-evaporation-mechanism}
		\end{figure}
		\par
		\autoref{fig5:schematic-evaporation-mechanism}(b), shows the various stages of evaporation for a microemulsion droplet. During Stage II, preferential vaporization occurs between the two volatile components (oil and water), while the surfactant AOT does not evaporate due to its non-volatile nature. Thus, the nature of base oil and its properties significantly influence the evaporation rate for a microemulsion droplet. Although the dispersed water sub-droplets can contribute to the overall evaporation rate, their influence is contingent upon transport to the droplet surface. In essence, the rate of evaporation in microemulsion droplets is dictated by the properties of base oil, and the dispersed water sub-droplets must traverse to the surface to participate in evaporation.
		\par
		\par
		Upon heating a microemulsion droplet, the temperature rise induces an increase in the interfacial tension at the interface between the base oil and dispersed water sub-droplets. This elevated temperature could lead to the detachment of AOT molecules from the interface, exceeding a critical threshold. Furthermore, evaporation of liquid mass from the droplet alters the proportion of the volatile component \cite{friberg1993evaporation}. Given the high sensitivity of microemulsions to compositional changes, it is likely that the microemulsion deviates from its original composition and structure, determined using scattering techniques in a closed cuvette. Understanding how the nanostructure and transport properties of these thermodynamically stable microemulsions evolve under elevated temperature and evaporative mass loss remains a significant challenge experimentally. These detached AOT molecules may then aggregate within the droplet to form larger AOT particles (see \autoref{fig5:schematic-evaporation-mechanism}). For a given $\omega$, the water and AOT volume fraction increases with the increase in $\phi$. This increased concentration of AOT molecules inherently increases the probability of their agglomeration. Consequently, microemulsions with higher $\phi$ values can be expected to exhibit a greater number of agglomerated AOT particles compared to those with lower $\phi$.
		\par
		The overall evaporation rate of a microemulsion droplet is anticipated to be influenced by its constituents and properties, contingent on irradiation intensity. While the readily available base oil at the droplet surface is the primary determinant, dispersed water can also contribute to evaporation when transported to the surface. There are three different mechanisms through which the water sub-droplets may get exposed to the surface; (1) receding droplet surface exposing water sub-droplets due to evaporation, (2) diffusive transport, and (3) convective transport due to internal flow. The receding time scale of the droplet surface is the same as the evaporation time scale, which is of the order $t_\text{evp} \sim O(10^1)$ at $I^\ast = 0.05$ and $\sim O(10^0)$ at $I^\ast = 0.1$. The diffusion time scale ($t_\text{dif}$) can be calculated using the $t_\text{dif} \sim D_0^2/\mathcal{D}_\text{eff}$, where ${D}_\text{eff}$ is the effective diffusivity of nanometer-sized dispersed water sub-droplets. The diffusivity of these small sub-droplets is extremely low ($\sim O(10^{-11})$)\cite{rastogi_diesel_2019}, therefore the diffusion time scales are very large $t_\text{dif} \sim O(10^3)$ (for decane microemulsion at $\omega=10, \phi=0.1$) compared to evaporation time scale. The convection time scale ($t_\text{cov} \sim D_0/u_\text{cov}$), where $u_\text{cov}$ is the velocity inside the acoustically levitated droplet. The convection velocity can be calculated using $u_\text{cov} = \sqrt{\dfrac{\mu_\text{a}\rho_\text{a}}{\mu_\text{d}\rho_\text{d}}} u_\text{max}$, where, $\mu_\text{a}$ and $\rho_\text{a}$ and dynamic viscosity and density of the air medium surrounding the droplet, and, $\mu_\text{d}$ and $\rho_\text{d}$ and dynamic viscosity and density of the droplet\cite{yarin_evaporation_1999}. The $u_\text{max}$ is the maximum acoustic streaming velocity which is given as $u_\text{max} = c_0Ma$, where $c_0$ is the speed of sound and $Ma$ is the Mach number, which is related to the acoustic sound pressure level ($SPL$), as $SPL = 197+20\log({Ma})$\cite{saha_velocity_2012,yarin_evaporation_1999}. The convection velocity due to internal flow inside the droplet is $u_\text{cov}\sim 250 \text{ mm/sec}$ which gives the convective time scale of $t_\text{cov} \sim O(10^{-3})$. Although the convective time scale is much smaller compared to the evaporation time scale, it is important to note that the streamlines within an acoustically levitated droplet are azimuthal rather than radial. Therefore, it can be assumed that the convective velocity does not influence water sub-droplet transport towards the surface. As a result, diffusive and convective transport can be assumed to have minimal influence on the overall evaporation of the microemulsion droplet. Therefore, the evaporation of water sub-droplets primarily depend on the receding of evaporating droplet surface. A schematic diagram illustrating the role of different components in microemulsion droplet evaporation is presented in \autoref{fig5:schematic-evaporation-mechanism}(c), however, it is important to note that experimental observation of this phenomenon is extremely difficult due to nanometer size ($\sim 5-30$ nm\cite{rastogi_diesel_2019}) of these water sub-droplets.
		\begin{figure}
			{\includegraphics[width=0.95\textwidth]{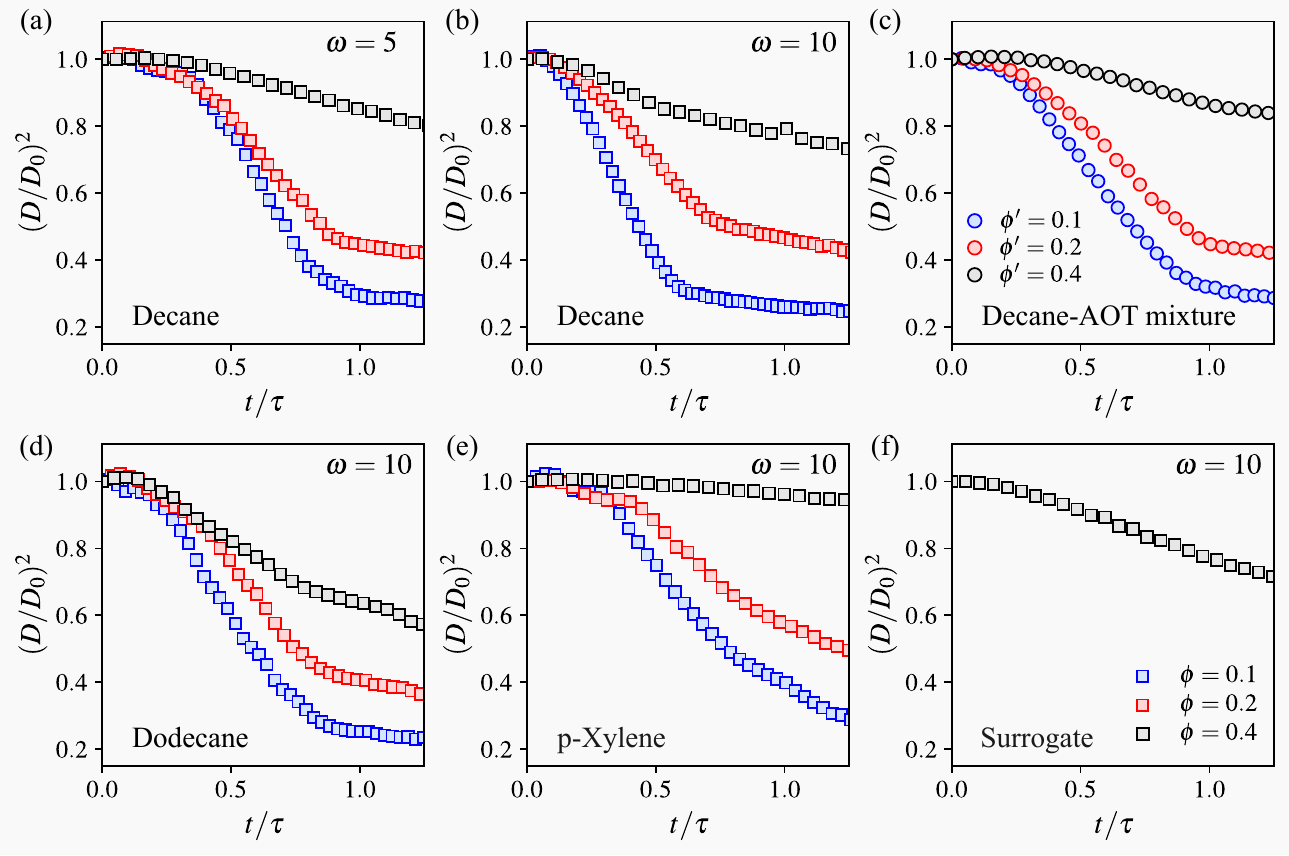}}
			\caption{Comparing the temporal evolution of non-dimensional droplet diameter for different values of $\phi$ at (a) $\omega = 5$ (b) $\omega = 10 $ for decane microemulsion. (c) depict the droplet diameter regression for the mixture of decane and surfactant (without water, $\omega=0$) revealing a similar trend of evaporation for different $\phi$, here $\phi$ solely represents the volume fraction of AOT within the decane-AOT solution. Figures (d),(e), and (f) illustrate the temporal evolution of droplet diameter for dodecane, xylene, and surrogate microemulsion droplets, respectively. All plots shown are at a constant laser irradiation intensity of $I^\ast = 0.05$.}
			\label{fig6:regression-all-p5}
		\end{figure}
		The temporal evolution of normalized droplet diameter is shown in the \autoref{fig6:regression-all-p5}, and it is evident that a strong correlation exists between the dispersed phase volume fraction ($\phi$) and the evaporation rate of microemulsion droplets. For all the values of $\omega$ (water to AOT molar ratio), as $\phi$ (dispersed phase volume fraction) increases, the evaporation rate decreases across all the base oils investigated in this study. The decrease in evaporation rate is primarily governed by two key factors: (1) lower oil volume fraction ($\phi_\text{oil}$), (2) increase in the number of water sub-droplets present in the suspended microemulsion droplet when $\phi$ increases. As discussed before, vaporization at the surface of the droplet is the combination of continuous vaporization of bulk liquid (base oil) and intermittent vaporization of water sub-droplets once exposed at the surface). As $\phi$ increases, the corresponding volume percentage of the base oil in the microemulsion droplet decreases (the volume percent of different components in a microemulsion droplet is tabulated in the \textcolor{blue}{Table S1} for $\omega=$10 and 40). For instance, a droplet with $\phi= 0.1$ signifies a 90\% oil composition by volume, with the remaining 10\% constituting the dispersed phase (comprises of water and AOT dictated by $\omega$ (see \autoref{eqn:phi-omega-relations})). The abundant base oil readily replenishes at the droplet surface, leading to a higher evaporation rate, particularly for a lower value of $\phi$. To isolate the effect of non-volatile surfactant on evaporation, samples with an oil-AOT mixture were also prepared without the addition of water. The oil-AOT mixtures were prepared by adding an exact amount of AOT into oil, corresponding to the microemulsion ($\omega=10$). \autoref{fig6:regression-all-p5}(c) shows the droplet lifetime history for decane-AOT mixture ($\omega^\prime=10$), for different values of $\phi^\prime$. It is observed that an increase in $\phi^\prime$ results in a decrease in the evaporation rate of the droplet (oil-AOT). Here $\omega^\prime=10$ represents the decane-AOT mixture corresponding to decane microemulsions at $\omega=10$ and $\phi^\prime$ indicates the corresponding $\phi$ of the microemulsion ($\omega=10$) which contains the same amount of AOT used to prepare the microemulsion. The prime ($^\prime$) notation has been adapted in this work to distinguish oil-surfactant mixture from microemulsions.
		\par
		While the evaporation process is primarily driven by the nature of the base oil, the evaporation of water sub-droplets is contingent upon their exposure to the surface. As time progresses, the $\phi_\text{AOT}$ continuously increases because of a decrease in the volume of volatile components. This effect becomes even more pronounced for larger $\phi$ due to the high initial concentration of AOT.
		\begin{figure}
			\centering
			\includegraphics[width=0.8\textwidth]{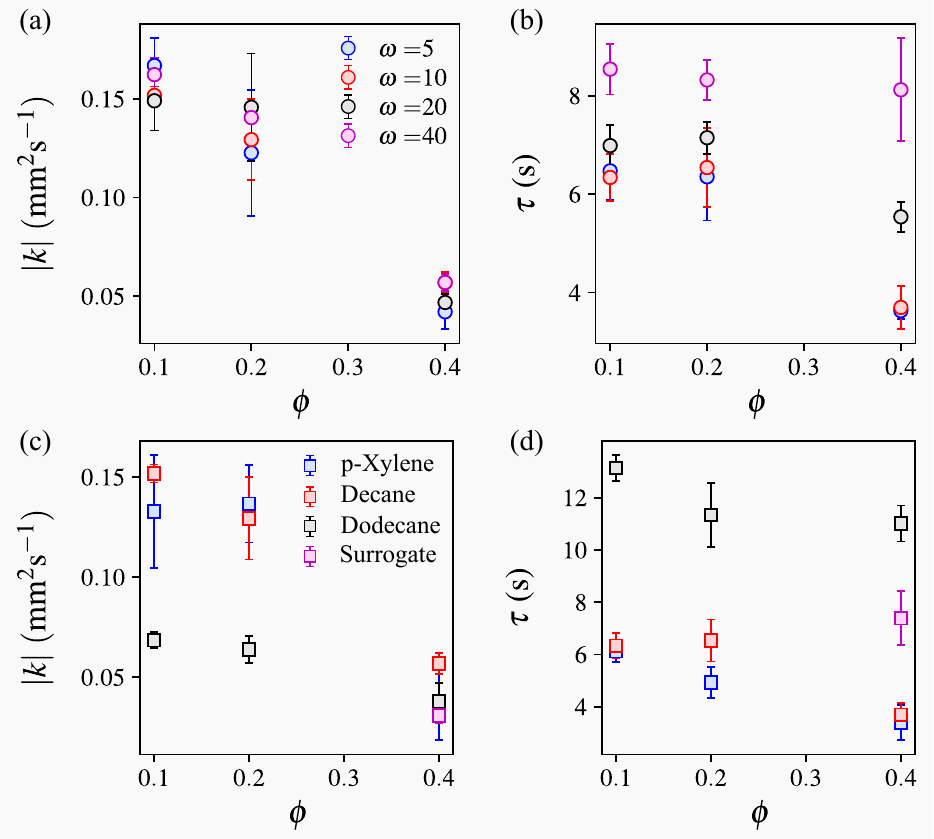}
			\caption{(a) and (c) Shows the droplet evaporation rate ($|k|$), as a function of $\phi$ for different $\omega$ and base oils, respectively. (b) and (d) Shows the droplet lifetime ($\tau$), as a function of $\phi$ for different $\omega$ and base oils, respectively at $I^\ast = 0.05$.   }
			\label{fig7:evaparation_rate_time_all_p5}
		\end{figure}
		In general, the overall droplet evaporation rate ($k$) can be estimated using the equation\cite{mehrizi_evaporation_2022}, 
		\begin{equation}
			k = \frac{D(\tau)^2 - D_0^2}{\tau}
		\end{equation}
		where, $\tau$ is the total evaporation time, $D(\tau)$ and $D_0$ are the final and initial droplet diameter, respectively. In microemulsions, to estimate the steady evaporation rate ($k$) from the experimental data, we have calculated the slope of the square of droplet diameter vs the time curve in the steady evaporation phase. The evaporation rate was determined by employing the least squares method implemented within a Python code, which yielded a minimum $R^2$ value of 0.98, indicating a highly linear correlation between the squared diameter of the droplet and time. The evaporation rate for decane microemulsion for different values of $\omega$ as a function of $\phi$ is plotted in the \autoref{fig7:evaparation_rate_time_all_p5}(a), which shows that the increase in $\phi$ reduces the rate of evaporation during Stage II. From \autoref{fig6:regression-all-p5} and \autoref{fig7:evaparation_rate_time_all_p5}(a), it is evident that the difference in evaporation rate between $\phi = 0.1$ and 0.2 is minimal, whereas a significant reduction occurs for $\phi = 0.4$. The significant decrease in evaporation rate can be attributed to the higher concentration of AOT. At higher $\phi$ (for example, $\phi$ = 0.4),  a denser packing of AOT aggregates at the droplet surface occurs due to a higher concentration of AOT, which reduces the effective surface area for evaporation.
		The evaporation lifetime ($\tau$) of the droplet is plotted in the \autoref{fig7:evaparation_rate_time_all_p5}(b) and (d) for decane microemulsion and different base oils as a function of $\phi$ at $I^\ast = 0.05$. It is seen that the droplet lifetime also decreases with an increase in $\phi$. A higher $\phi$ leads to the inhibition of evaporation due to the reasons explained above, which eventually terminates the evaporation process much earlier. In essence, the evaporation ends much earlier for higher $\phi$, resulting in a shorter evaporation lifetime.  
		\par
		To further understand the evaporation process in a microemulsion system, we formulate a parameter, $\eta$, defined as the ratio of volume fraction of volatile to non-volatile components of the microemulsion droplet. The $\eta$ can be expressed as
		\begin{equation}
			\eta = \frac{1 - \phi_\text{AOT}}{\phi_\text{AOT}}.
		\end{equation}
		As shown in the \autoref{fig7:evaparation_rate_time_all_p5}(a) and (c), as the value of $\eta$ increases, the evaporation rate of the droplet also increases. This is expected because increasing $\eta$ essentially indicates an increase of volatile components in the droplet and a reduction in the non-volatile AOT, leading to enhanced evaporation, as discussed before. This behavior is consistent for all the microemulsions with different base oils.
		\begin{figure}
			\centering
			\includegraphics[width=0.85\textwidth]{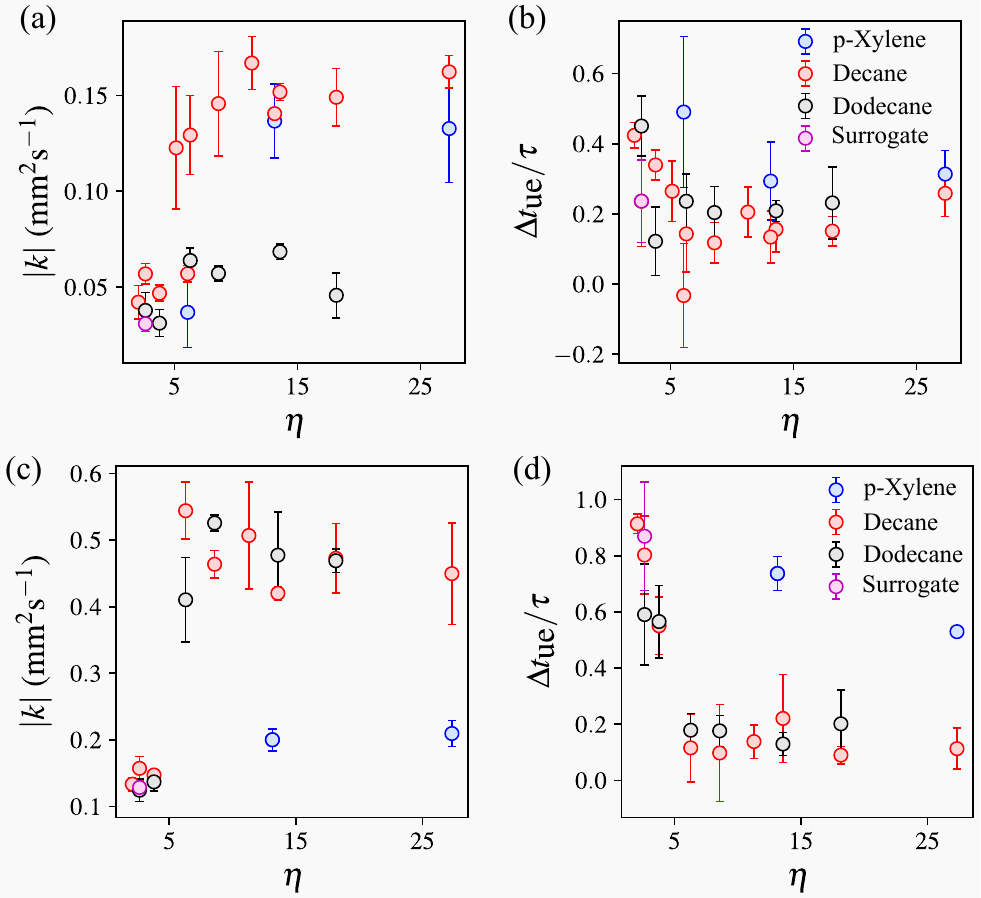}
			\caption{(a) and (d) Shows the droplet evaporation rate ($|k|$), as a function of $\eta$ for different base oils at $I^\ast = 0.05$ and 0.1 respectively. (b) and (d) Shows that the variation of normalized evaporation time during the unsteady $\Delta t_\text{ue}/\tau$ phase as a function of $\eta$, at $I^\ast = 0.05$ and 0.1 respectively}
			\label{fig8:evaporation_parameters_eta}
		\end{figure}
		
		\paragraph{Role of the base oils (continuous phase): }\label{prgphsec:role-base-oils}
		So far, we discussed the effect of $\omega$ and $\phi$ on the evaporation characteristics. \autoref{fig7:evaparation_rate_time_all_p5}(c) shows the evaporation rate of microemulsions ($\omega = 10$) having different base oil as a function of $\phi$ at $I^\ast = 0.05$. Decane and xylene show similar values of evaporation rate ($k$). However, the dodecane microemulsion exhibits a much lower evaporation rate for all the values of $\phi$ except at $\phi=0.4$. The evaporation characteristics of the microemulsions having different base oils can be partially explained based on the vapor pressure ($p_v$) of base oil. The vapor pressure of dodecane, decane, and xylene are of the order $O (10^{-2})$, $O (10^{-1})$ and $O (10^{1})$ respectively, as shown in the \autoref{tab1:base-liquid-properties}. As discussed earlier, the characteristics of the base oil primarily control evaporation, and since the vapor pressure of dodecane is one order lower than decane, the dodecane shows a much lower evaporation rate compared to the decane microemulsion. However, despite xylene having the highest vapor pressure among all the base oils, the xylene microemulsion shows an evaporation rate similar to that of decane microemulsion at $I^\ast = 0.05$. The evaporation of a microemulsion droplet is a complex process not solely governed by the characteristics of base oil. Additional factors, such surfactant agglomeration, possibly the nanostructure, can significantly influence the overall evaporation rate.
		The relatively reduced evaporation rate of the xylene microemulsion can be attributed to the interplay between the vapor pressure of base oil and the agglomeration of surfactant. Due to its high vapor pressure, xylene initially evaporates rapidly from the droplet surface. This rapid evaporation leads to an early and significant rise in the relative concentration of non-volatile AOT at the surface. These particles lack sufficient time to diffuse and achieve a homogeneous distribution throughout the droplet. The high concentration of AOT slows the transport of liquid components toward the droplet surface, effectively limiting the evaporation rate of xylene by reducing the effective surface area for evaporation. It is important to distinguish this Stage from a phase where the droplet surface is entirely covered with AOT particles (which comes much later), effectively halting evaporation.
		\par
		\autoref{fig8:evaporation_parameters_eta}(a) and (c) shows the evaporation rate ($k$) for different base oils as a function of $\eta$ at $I^\ast=0.1$. It is observed that, at a high heating rate, the evaporation rate of the decane and dodecane microemulsion is similar, and the xylene microemulsion exhibits an even lower evaporation rate. This indicates that the combined influence of the vapor pressure and the elevated concentration of AOT near the droplet surface becomes increasingly significant at higher heating rates. This indicates that the combined influence of the vapor pressure and the elevated concentration of AOT particles near the droplet surface becomes increasingly significant at higher heating rates. It is to be noted that this effect is also expected to occur for the emulsion with higher $\phi$. It is also shown in the \autoref{fig8:evaporation_parameters_eta}(a), and (c) that, for the lower value of $\eta$, we observe very closely packed data points for all the base oils, indicating similar evaporation rates irrespective of the heating rate. Additionally, the evaporation time also depends on the volatility of the base oils and the relative concentration of the AOT. As shown in \autoref{fig7:evaparation_rate_time_all_p5}(d), the dodecane microemulsion exhibits the highest evaporation time followed by surrogate, decane, and xylene. This is also consistent with the trend in their evaporation rate.
		
	\subsubsection{Stage III : Unsteady evaporation  and shell formation}\label{sec3-2-3:stage-III}
	During Stage II, the microemulsion droplet exhibits a constant rate of evaporation, and as time progresses, this process eventually transitions to a stage where the rate of diameter reduction progressively diminishes until it becomes negligible or constant ($\sim0$), as shown in \autoref{fig6:regression-all-p5} ($t/\tau = 1$ indicates the end of Stage III). Between the end of Stage II and the point where the evaporation rate approaches zero is called \emph{unsteady evaporation} phase (i.e., Stage III). Most of the liquid components evaporate during Stage II. As a result, the relative concentration of surfactant inside the droplet increases drastically at this Stage (II). This creates a liquid-deficient droplet with high AOT concentration. As time progresses, the transport of liquid components to the surface becomes even more difficult. Because of these two phenomena, the evaporation rate of the microemulsion droplet decreases continuously until the evaporation completely stops. \autoref{fig8:evaporation_parameters_eta}(b) and (d) show the evaporation time during the unsteady evaporation phase of the droplet. It is observed that the droplet with a lower value of $\eta$ exhibits a larger evaporation time ($\Delta t_\text{ue}$). This behavior can be attributed to the higher concentration of the AOT relative to the concentration of volatile components (oil and water) leading to an early transition from steady to unsteady evaporation (see \autoref{fig6:regression-all-p5}). As a result, the majority of the liquid evaporation occurs during Stage III and at a slower rate, which leads to an increase in unsteady evaporation time scale ($\Delta t_\text{ue}$).
	\par
	Unlike pure liquids, complete evaporation of the droplet does not occur, instead a residual material, often referred to as a ``shell" remains following the evaporation process. The basis of the formation of this spherical solid shell can be attributed to the rapid packing of agglomerated AOT particles at the droplet surface compared to their diffusion to the bulk\cite{raju_controlling_2018,ahmed_2020}.
	\begin{figure}
		\centering
		\includegraphics[width=0.75\textwidth]{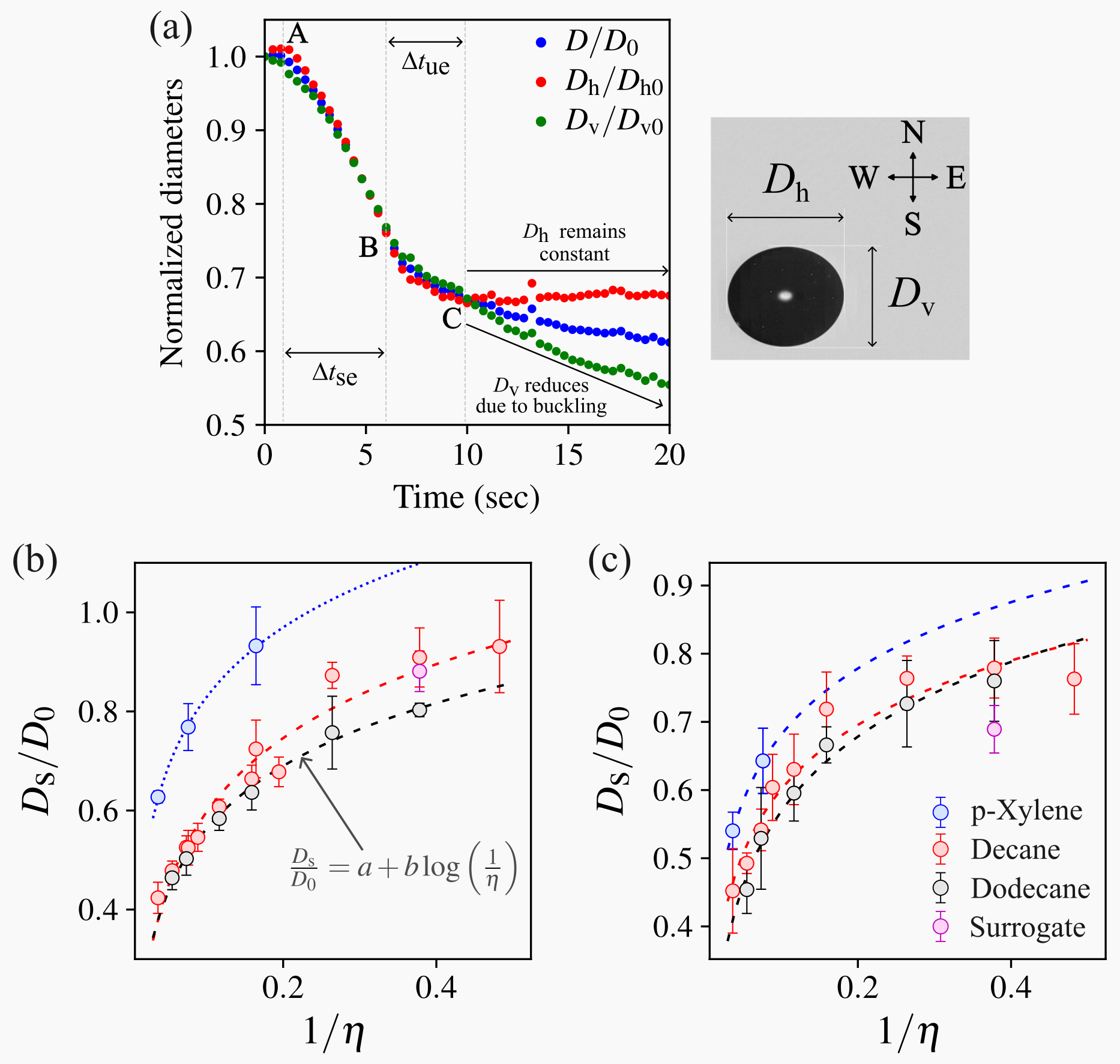}
		\caption{(a) Plotted the variation of horizontal ($D_\text{h}$) and vertical ($D_\text{v}$) diameters of the decane microemulsion ($\omega=10, \phi =0.2$) droplet with the equivalent droplet diameter as a function of time ($t$). (b) and (c) Shows the relationship between residual shell size and AOT volume fraction ($\phi_\text{AOT}$) for microemulsions with different base oils at $I^\ast = $ 0.05 and 0.1, respectively. Here, the shell size $D_\text{s}$ is plotted against the parameter $\eta$, which represents the relative volume fraction of the volatile to a non-volatile component of the microemulsion.}
		\label{fig9:decane-dhdv-residue-size}
	\end{figure} 
	The formation of this spherical shell suggests that a significant amount of the microemulsion components continue to remain even after the evaporation stops. This un-evaporated mass of the droplet gets encapsulated by a layer of agglomerated surfactant, completely closing the shell and thus preventing further mass transfer from the droplet. However, this residue formed post-evaporation is not unique to the microemulsion droplets. These shell-like structures have been observed for several multi-component droplets containing non-volatile components, for instance, nano-particle laden droplets\cite{mehrizi_evaporation_2022,saha_2010}, polymeric droplets\cite{gannena_bubble_2022} and even in biological fluids such as respiratory droplets\cite{vejerano_physico-chemical_2018, seyfert_stability_2022}. While in all such cases, the evaporating droplet results in residue formation, the characteristics of this residue differ significantly. For instance, respiratory droplets do not undergo buckling (a phenomenon observed in the present work, which will be discussed further) but instead show crystal formation\cite{chaudhuri_2020}.
	\par  
	At the onset of the shell formation, droplet evaporation stops, which translates to a negligible change in the size of the residual structure.  Therefore, this should indicate a zero $\left(\dfrac{d}{dt}D(t)^2\sim 0\right)$ slope in the diameter vs time curve (see \autoref{fig3:pure_vs_me_droplet_evaporation}(c), \autoref{fig6:regression-all-p5}). As evident from the figures, it is difficult to ascertain the point where the droplet evaporation ends. We observe a persistent reduction in droplet size $\left(D(t)^2/D_0^2\right)$ for a significant time interval (10 - 30 seconds), even after the shell formation. To elucidate this unexpected observation, a detailed analysis of droplet diameter evolution is carried out. \autoref{fig9:decane-dhdv-residue-size} (a) presents the temporal evolution of horizontal ($D_\text{h}$) and vertical ($D_\text{v}$) characteristic lengths (diameter) of the microemulsion droplet along with the equivalent droplet diameter ($D$). It is observed that, during stages II and III, the horizontal and vertical diameters reduce proportionally with the equivalent diameter. For an evaporating droplet, it is expected that both of its characteristic diameters (horizontal and vertical) should decrease with time as a result of a reduction in the overall droplet volume.
	However, a significant deviation from this anticipated behavior of the droplet diameters has been observed after point \textbf{C}, as shown in \autoref{fig9:decane-dhdv-residue-size}(a).  While the horizontal diameter appears to reach a plateau, indicating minimal change in droplet size with time, the vertical diameter continues to decrease in size. This discrepancy between the horizontal and vertical diameter variations is attributed to the \emph{buckling} of the spherical shell. The evaporation is considered to stop at the point (\textbf{C}), where the horizontal diameter becomes constant.
	\begin{figure}[ht]
		\centering
		\includegraphics[width=0.95\textwidth]{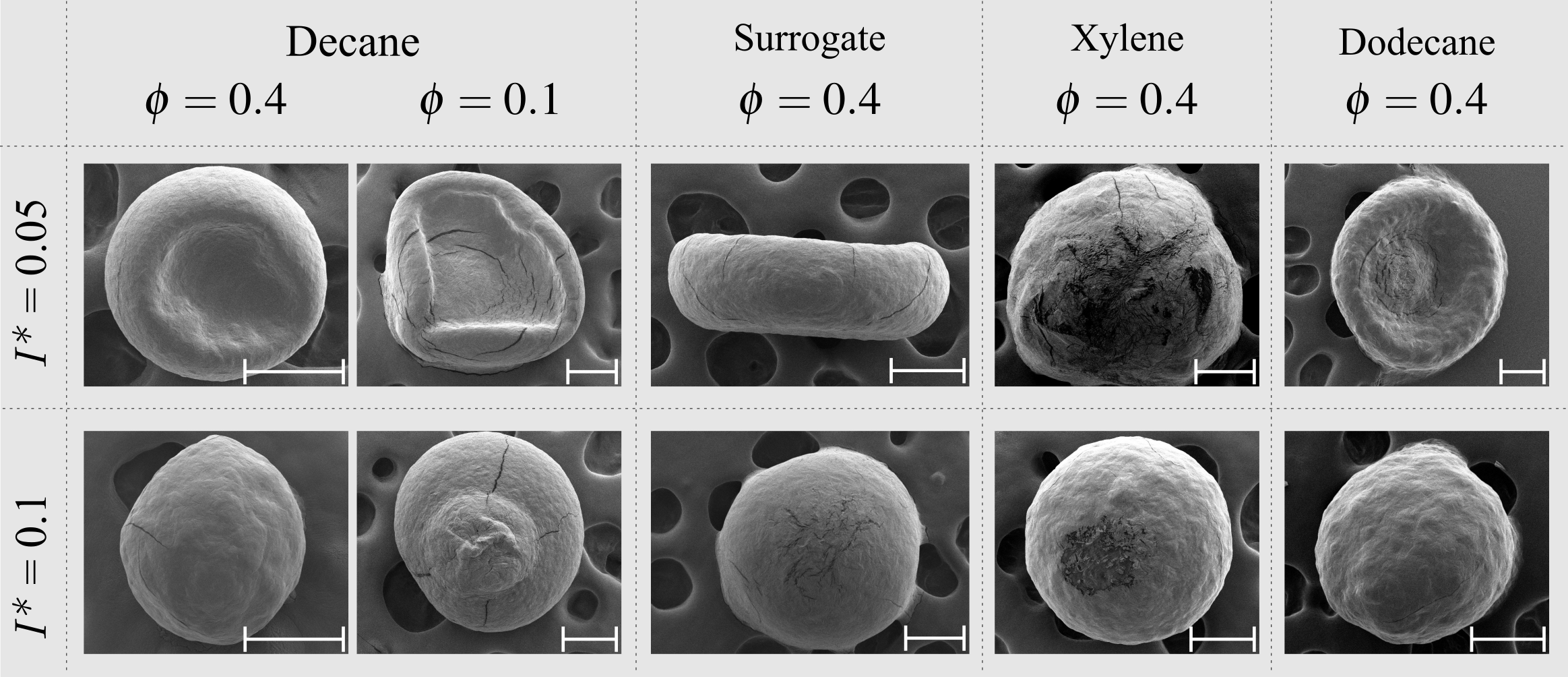}
		\caption{SEM images of the residue are shown for different values of $\phi$ and irradiation ($I^\ast$) at $\omega = 10$. The images show buckling of the spherical shell at low irradiance($I^\ast = 0.05$); however, for $I^\ast = 0.1$, the shell shows spherical morphology for the majority of the cases (All the scale bars represent a length of 200 microns).}
		\label{fig:fig10:SEM_images_shell}
	\end{figure}
	\par
	The spherical shell exhibits localized buckling, majorly at its north (\textbf{N}) and south poles (\textbf{S}). The deformation of these shells is typically attributed to capillary forces. Liquid evaporation at the droplet surface induces the formation of microscopic pores within the AOT shell. These pores function as tiny menisci, where the Laplace pressure ($\delta p \sim 2\sigma/r_\text{pore}$) acting just inside the meniscus falls below the ambient atmospheric pressure. The curvature-induced capillary force pulls the liquid out of the pores and pulls the spherical shell inwards, eventually leading to the buckling of the shell\cite{tsapis_onset_2005}.
	In addition to capillary forces, acoustic radiation forces acting on the droplet might also influence the observed shell buckling phenomenon. It is observed that the buckling primarily occurs at the north (\textbf{N}) and south (\textbf{S}) poles of the shell, which are also seen in the SEM images of the residual shell (\autoref{fig:fig10:SEM_images_shell} at $I^\ast = 0.05$). This observation suggests a potential role of acoustic pressure in the buckling phenomenon, particularly since the droplet is subjected to a continuous acoustic field. Notably, the spherical shell may not always undergo buckling. A spherical shell morphology can also be observed at the end of Stage III, which is referred to as \emph{shell jamming} (see \autoref{fig:fig10:SEM_images_shell} at $I^\ast = 0.1$) xylene microemulsion at $I^\ast = 0.1$). It is observed that initially, the north pole of the shell buckles, which is followed by buckling in the south pole (see \autoref{fig3:pure_vs_me_droplet_evaporation}(b)). The observed behavior can be attributed to the influence of gravity. During evaporation, the liquid within the droplet may tend to accumulate at the south pole due to gravitational pull and surface tension forces acting at the interface ensure the spherical shape.
	\par
	The final diameter of the residual spherical shell at the end of Stage III is measured from the shadowgraph images. The shell size depends on both the evaporation rate and the volume fraction ratio ($\eta$) of the volatile to non-volatile components within the parent droplet. \autoref{fig9:decane-dhdv-residue-size}(b) and (c) show the variation of the shell size ($D_\text{s}$) as a function of $\eta$ at $I^\ast = 0.05$ and $0.1$, respectively. It can be seen that a higher shell size $D_\text{s}$ is inversely proportional to the volume fraction ratio ($D_\text{s} \propto 1/\eta$). The shell size of the xylene microemulsion exhibits large shell sizes compared to the other microemulsions. The larger shell size can be attributed to the high vapor pressure of xylene which leads to its faster evaporation. Therefore, the rate of accumulation of AOT particles is much faster than their diffusion/transport throughout the droplet causing an early and larger shell formation.
	Typically, for xylene microemulsion, no buckling has been observed at a higher heating rate (see \autoref{fig:fig10:SEM_images_shell}). The faster evaporation rate of xylene causes the shell to form early such that the solid residual structure can sustain deformation due to the acoustic radiation pressure.
	
	A smooth temporal variation of non-dimensional droplet diameter for all microemulsions has been observed at IR irradiation intensity of $I^*=0.05$. Similar behavior has been observed at $I^\ast=0.1$ as well, except for xylene and decane microemulsion. The xylene ($\phi=0.4$, $\omega=10$) and decane ($\phi=0.2,0.4$, $\omega=40$) microemulsions show bubble nucleation and breakup at $I^*=0.1$. Note that the focus of the present study is to investigate the evaporation characteristics in stable microemulsion droplets. Thus, the study of bubble nucleation and breakup lies beyond the scope of this work and is not addressed herein.
	
\section{Conclusions}\label{sec4:conclusions}
We explore the evaporation dynamics of stable microemulsion droplets in a contactless environment using an infrared continuous laser. The key evaporation characteristics of stable emulsion droplets are as follows.
\begin{enumerate}
	\item The microemulsion droplet exhibits three distinct stages of evaporation: pre-heating, steady evaporation, and unsteady evaporation. At the end of evaporation, the droplets result in a spherical residue.
	\item Pre-heating corresponds to the transient heating period of the droplet, where its surface temperature increases and becomes constant after a certain time period, i.e., preheating time ($\Delta t_\text{ph})$. The ranges from $\Delta t_\text{ph} \sim 1.72 $ seconds at irradiation intensity of $I^\ast = 0.05$, whereas it reduces to $\sim 0.76$ seconds at $I^\ast = 0.1$.
	\item During Stage II, the droplet undergoes steady evaporation, and the rate of evaporation is observed to decrease with the increase in $\phi$ for a constant $\omega$. It is shown that the evaporation of a microemulsion droplet is governed by the complex interplay between its constituents and their properties. A parameter $\eta$ was introduced, which indicates the cumulative influence of various factors ($\phi, \omega$) affecting the evaporation process. Moreover, the nanostructure of microemulsion may have a role in droplet evaporation, however, more evidence is needed to confirm its influence on evaporation.
	\item As time progresses, the evaporation rate continuously decreases until the evaporation stops, (Stage III). Finally, the droplet transforms into a solid residual structure. The morphology of this residual shell is spherical, which further undergoes buckling at its south and north poles. This shell size $D_\text{s}$) is observed to increase as the value of $\eta$ decreases ($D_\text{s} \propto \eta^{-1}$).
	\item Among all the base oils used in this study, the xylene microemulsion shows a larger shell size in comparison to other microemulsions with a different oil phase. This may be attributed to its high vapor pressure of the xylene.
	\item At higher heating, we usually observe a nearly spherical shell in contrast to a buckled non-spherical structure for a lower heating rate.
\end{enumerate}

\section{Author's Contribution}
\textbf{Bal Krishan:} Experiments, Data curation, Data Processing and  Analysis, Writing – original manuscript, Writing – review \& editing.
\textbf{Preetika Rastogi:} Experiments, Data curation, Scanning Electron Microscopy, Conceptualization, Writing – review \& editing.
\textbf{D. Chaitanya K. Rao:} Experiments, Conceptualization, Theoretical Analysis, Writing – review \& editing.
\textbf{Niket S. Kaisare:} Conceptualization, Resources, Writing – review \& editing, travel grant.
\textbf{Madivala G. Basavaraj:} Conceptualization, Resources, Writing – review \& editing, travel grant.
\textbf{Saptarshi Basu:} Conceptualization, Resources, Theoretical Analysis, Writing – review \& editing.
	
\begin{acknowledgement}

The authors thank Dr. Durbar Roy and Vadlamudi Gautham for insightful discussions and suggestions, Dr. Gannena K.S. Raghuram for his assistance with the rheology experiments and G. Prashanth for help with imaging on the high resolution scanning electron microscopy (HR-SEM) procured through a DST-FIST grant at the Department of Chemical Engineering, IIT-Madras. We also acknowledge Dr. Najet Mahmoudi for SAXS and WAXS experiments carried out at the ISIS Neutron and Muon Source at the Rutherford Appleton Laboratory, Didcot, UK. Funding from SPARC is gratefully acknowledged.

\end{acknowledgement}

\vspace{10em}
\noindent\fontsize{16}{16}{\textbf{Data Availability Statement}}\\ 
The data that support the findings of this study are available
from the corresponding author upon reasonable request.

\vspace{2em}
\noindent\fontsize{16}{16}{\textbf{Author Declarations}}\\ 
The authors have no conflicts to disclose.

\bibliography{references}

\end{document}